# Optimizing coverage of 3D Wireless Multimedia Sensor Networks by means of deploying redundant sensors

Authors Name/s per 1st Affiliation *(Shasha.Xu)*
Tianjin University
Tianjin, China
shasha.xu91@foxmail.com

Authors Name/s per 2nd Affiliation *(Weijie. Lyu)*
Tianjin University
Tianjin, China
weijielv@tju.edu.cn

Authors Name/s per 3rd Affiliation *(Hulin.LI)*
Xi'an Polytechnic University
Xi'an, China
serenelly_lee@163.com

*Abstract*—Coverage is one of the fundamental issues in wireless multimedia sensor networks (WMSNs). It reflects the ability of WMSNs to detect the fields. Motivated by the existing-enhancing algorithm of traditional 2D WMSNs, a new 3D WMSNs sensing model is established and a new coverage-enhancing algorithm based on this model is proposed. This algorithm defines the sensing model as trapezoidal pyramid, calculates the key parameters (tilt angle) then improves coverage ratio by optimizing it. However, there still exists redundant sensors in this optimized networks. Aiming at efficiently utilizing these redundant sensors and enhancing coverage ratio, the authors selects the redundant sensors by introducing the set cover model algorithm, further deploys them to the uncovered area following the greedy policy, so that the whole path coverage performance of WMSNs is enhanced.

*Keywords—wireless multimedia sensor networks, 3D sensing model, coverage-enhancing, redundant nodes optimization*

## I. INTRODUCTION

Due to many attractive characteristics of sensor nodes such as small size and low cost, wireless sensor networks (WSNs) have become adopted to many military and civil applications including military surveillance, smart homes, remote environment monitoring, and in-plant robotic control and guidance [1-3]. Wireless multimedia sensor network (WMSNs), an advancing form of wireless sensor network (WSNs), is distributed sensing networks consisted of video cameras that have sector sense area and restricted by a finite field of view, which may contains several sensing directions [4-5]. One of the most important issues in WMSNs is sensing coverage, which reflects how well a sensor network is monitored or tracked by sensors. Perfected coverage is critical to sense target area and collect useful data [6]. Conventional researches concentrate on the coverage problems based on the 2D sensing model [2]. While the existing 2D sensing-model based on frame works are analytically more tractable in designing and evaluating the directional sensor networks for coverage-control, they cannot accurately characterize the practical application scenes. In addition, the 2D based schemes cannot be easily extended to address the coverage issues in 3D sensing model [7-10]. Hence, research about coverage problem in 3D WMSNs has much more practical significance. However, due to the high complexity in the design and analysis imposed by the 3D sensing model, most existing works focus on the simplified 2D sensing model [12]. To remedy this deficiency, Ji et al. have proposed a 3D directional sensor coverage-control model with tunable orientations and a greedy iteration based area coverage-enhance scheduling algorithm named GA-ACE [7]. Xiao et al. proposed another coverage-enhancing algorithm named TDPCA, the 3D sensing model is dived into pitch angle and deviation angle, then swarm optimization is used to eliminate overlap and blind area [11].

On this basis, we present a more sophisticated coverage-enhancing algorithm in this paper. The proposed algorithm defines the sensing model as trapezoidal pyramid. To maximize the area coverage is to maximize the area of the trapezoidal pyramid projected in the monitored scene plane during the process of slicing. For a more intuitive idea, we set up the formula of calculating the volume which the key parameters can be calculated, besides the tilt angle γ (vertical offset angle of the main sensing direction) can be optimized. However, there exists redundant sensors which contributes to the overlap area instead of the blind area in this optimized networks. The set covering model algorithm is applied to select these redundant sensor nodes and the greedy policy is followed to move them to the blind area. Thus the whole path coverage performance of the WMSN$_S$ is enhanced.

## II. 3D WMSN$_S$ SENSING MODEL

Different from traditional 2D sector model, 3D WMSN$_S$ sensing model depends on the characteristics of a Pan Tilt Zoom camera: the coverage area of sensor is constrained by the field of view, and functions as the projecting quadrilateral area in the monitored scene plane. In Fig. 1 (a), P-$V_1V_2V_3V_4$ is the rectangular pyramid with the monitored scene plane perpendicular to the sensing orientation. As presented in the Fig. 1 (b), when tilt angle $\gamma \neq 0$, the sensing model P-$D_1D_2D_3D_4$ functions as a trapezoidal pyramid with projecting trapezoid area □$D_1D_2D_3D_4$ in the monitored scene plane.

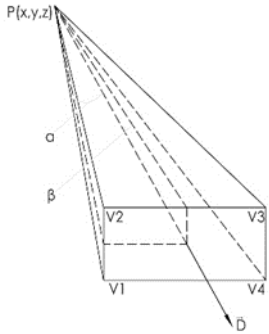

Fig. 1.(a) 3D WMSN$_S$ sensing model. (a) The rectangular pyramid with rectangular area □$V_1V_2V_3V_4$

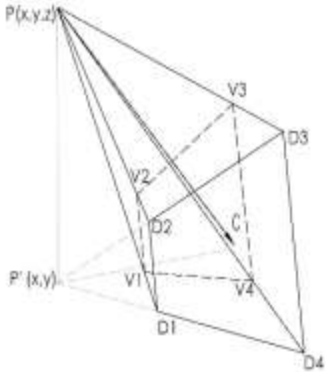

Fig. 1. (b) The trapezoidal pyramid with trapezoidal area □$D_1D_2D_3D_4$

*Definition 1:* 3D WMSN$_S$ sensing model can be described as 5-tuple $(P, \vec{C}, K, \alpha, \beta)$: P indicates the initial location (x, y, z) of the sensor in 3D space. $\vec{C} = (\gamma, \theta)$: $\gamma$ and $\theta$ indicates vertical and horizontal offset angle of the main sensing direction separately; K is the maximal value of $\gamma$ ($0 \leq \gamma \leq K$); The horizontal and vertical offset angles of the FOV around $\vec{C}$ are described as $\alpha$ and $\beta$, i.e., $2\alpha$ and $2\beta$ denoted the horizontal and the vertical FOV, respectively. Our 3D WMSN$_S$ sensing model with sensing orientation $\vec{C}$ is illustrated in Fig.2. From it we can see that if the tilt angle $\gamma$ is fixed, △$PD_1D_2$ slices the sector in horizontal plane.

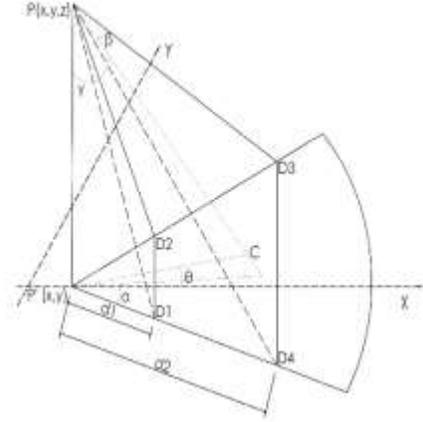

Fig. 2. The 3D WMSN$_S$ sensing model in the process of slicing

We utilize the X-Y-Z coordinate system to formulate our 3D sensing model. Considered as the centroid of sensing area, C is the intersecting point between the monitored scene plane that we define as z=0 in the X-Y-Z coordinate system and the central-line with the direction $\vec{C}$. The projection of P (x, y, z) is P' (x, y). Considering the horizontal component of the sensing direction $\vec{C}$, the projection of trapezoidal pyramid P-$D_1D_2D_3D_4$ is the traditional 2D sector with P' served as its vertex and $2\alpha$ served as its central angle. Considering the vertical constraint of the sensing direction $\vec{C}$, the triangle △$PD_1D_2$ moves along X axis, point $D_1$ turns into $D_4$, point $D_2$ turns into $D_3$, finally forms trapezoidal area □$D_1D_2D_3D_4$. And the angle between $PD_1D_2$ plane and $PD_3D_4$ plane is $2\beta$, additionally, line $\overline{PC}$ bisects this angle. In order to calculate the volume of trapezoidal pyramid, we introduce point $Q_1$ and point $Q_2$ marked as the segment of line $\overline{D_1D_2}$ and line $\overline{D_3D_4}$, respectively.

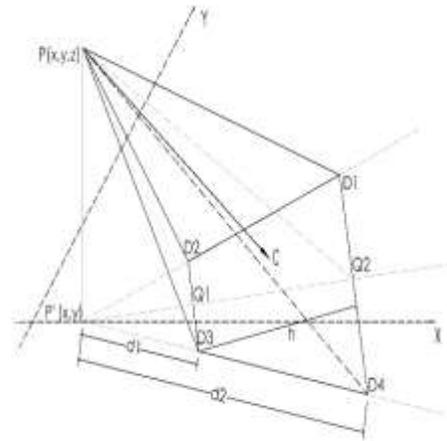

Fig.3 The 3D WMSN$_S$ sensing model with parameters

The main segments of the 3D directional sensing model used in the algorithm are calculated as follows.

$$|P'Q_1| = z \times tan(\gamma - \beta) \quad (1)$$

$$|P'Q_2| = z \times tan(\gamma + \beta) \quad (2)$$

$$d_1 = |P'Q_1|/\cos\alpha \quad (3)$$

$$d_2 = |P'Q_2|/\cos\alpha \quad (4)$$

The key parameters of trapezoidal pyramid are calculated as follows:

$$|D_1 D_2| = 2 \times |P'Q_1| \times \tan\alpha \quad (5)$$

$$|D_3 D_4| = 2 \times |P'Q_2| \times \tan\alpha \quad (6)$$

$$h = |P'Q_2| - |P'Q_1| \quad (7)$$

$$S = \frac{h}{2} \times (|D_1 D_2| + |D_3 D_4|) \quad (8)$$

$$V = S \times z \quad (9)$$

Collect all the formula and obtain the area S and the volume V of trapezoidal pyramid:

$$V = \frac{z}{2} \times (d_2 + d_1) \times (d_2 - d_1) \times \sin 2\alpha \quad (10)$$

Explanation: Increase the volume of trapezoidal pyramid can increase the coverage ratio, this formula shows the volume depends on the area of the trapezoidal which depends on horizontal FOV $2\alpha$ and length of z, $d_2$ ($d_1 \ll d_2$).

*Definition 2:* If 3D WMSN$_S$ sensing model can be denoted by $(P, \vec{C}, K, \alpha, \beta)$, and P' (x, y) is the projection of P(x, y, z) on the plane z=0. At any time, a point A on the monitored scene plane is said to be covered by a 3D WMSN$_S$ sensor node if and only if the following two conditions are met as shown in the Fig.4.
- $d_1 \leq |P'A| \leq d_2$, where point $A_1$, $A_2$ are intersection point between line $\overline{P'A}$ and $\overline{D_1 D_2}$ and $\overline{D_3 D_4}$.
- $\vec{P'A} \cdot \vec{C} \geq |\vec{P'A}| \cos\alpha$

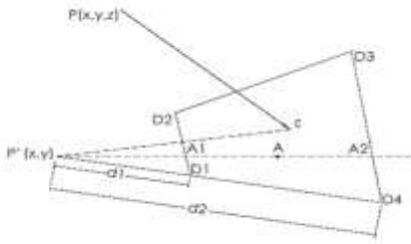

Fig. 4. Point P' in the projection sensing area

## III. PROBLEM FORMATION

As refereed in the above, the coverage area of sensor functions as the projecting quadrilateral area in the monitored scene plane.

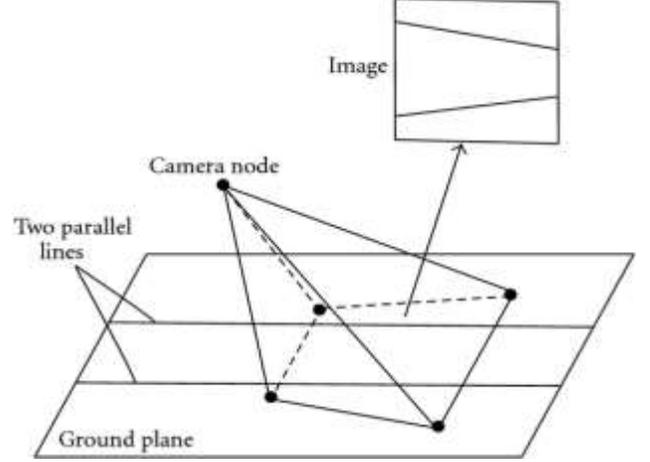

Fig. 5. The Schematic diagram of covergae problem in 3D WMSN$_S$

Explanation: camera node represents the 3D sensing mode, two parallel lines are introduced to prove that the projected sensing area is trapezoidal. The smaller picture named image shows that coverage problem in 3D has been successfully converted into 2D.

*Definition 3*: Deploy n sensor nodes, the coverage ratio η can be denoted as:

$$\eta = 1 - (1 - \frac{S}{G})^n \quad (11)$$

$$m = \ln(1 - \eta)/(\ln(G - S) - \ln G) \quad (12)$$

Explanation: as mentioned before, S represents the area of trapezoidal pyramid. G represents the monitored area. m sensor node can reach ratio η when the sensed area of the sensor node i is disjoint with that sensed by other sensor node.

To increase the coverage ratio η is to maintain the maximum area S, which can meet by m sensor nodes. Actually, there exists overlap area, so there would exist $(n - m)$ redundant sensor nodes in the 3D WMSN$_S$.

*Definition 4:* A grid is set to separate the monitored region to transmit continuous problem into discrete problem. A set of grids of the monitored region is denoted as Ω, and the set of grid sensed by at least one 3D WMSN$_S$ senor node is signified as $\Omega_C$:

$$\Omega_C = \cup\, \Omega_{ci}, 1 \leq i \leq n \quad (13)$$

$$F = \{\Omega_{c1}, \Omega_{c2}, ... \Omega_{cn}\} \quad (14)$$

Explanation: $\Omega_{ci}$ represents the set of discrete points that the ith sensor node covers, F represents the cover set of $\Omega_{ci}$, and the coverage ratio can be denoted in another way as follows:

$$\eta = \|\Omega_c/\Omega\| \quad (15)$$

In order to facilitate the analysis, we do some assumptions

for nodes and the monitored area as follows:

a. All randomly deployed 3D sensors are homogenous. Particularly, all sensors are 3D rotatable and able to control the change of its own sensing orientation.

b. Initial sensing orientation of all sensor nodes is randomly set and sensor nodes could move on XY plane after the initial deployment.

c. Each sensor knows its location information exactly. All sensors can communicate with each other.

Problem: For a group of sensors $(N_1, N_2, ... N_n)$ with initial random deployments, define $\eta'$ to be the one corresponding to the maximum trapezoid area sensed by every sensor node. Then how to meet $\eta'$ though adjusting tilt angle $\gamma$; how to select redundant sensor nodes which make no contribution to meet $\eta'$ from set F and how to relocate these redundant nodes to overlap and blind perception area effectively?

## IV. COVERAGE-ENHANCING BASED ON DEPLOYING REDUNDANT SENSORS

### A. Calcauting the best optimal tilt angle

Ideally, we control the length of $d_2$ to make $D_3D_4$ coincide with boundaries of sensing area, then the horizontal projection area is maximum. Actually, most sensor nods could not meet this ideal condition. According to the variation of tilt angle $\gamma$, divide the sensor nodes into four mutually disjoint $set_i$ i = (1,2,3,4). According to the formula that the volume depends on the area of trapezoid and the area of trapezoid depends on the length of d2. As shown in figure5, d$_2$ attains its maximum when the circle ○P'D₃D₄ is tangent to the four sides of rectangular area.

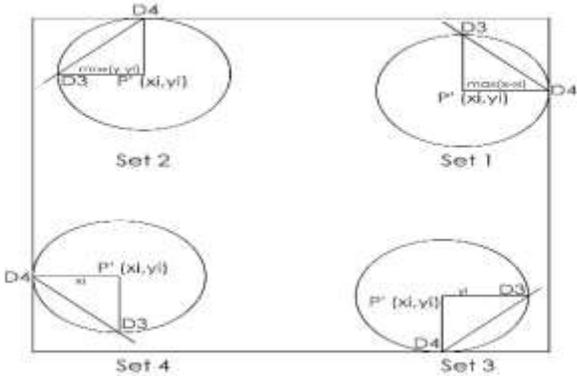

Fig. 6. Maximum area in the cutting process

Define that the maximum horizontal axis as $maxx$, the maximum vertical axis as $maxy$. Adjust tilt angle $\gamma$ with following formulas so as to create the maximum trapezoid area.

(1) if $i \in set_1$, then

$$\gamma(i) = \arctan((maxx - x(i))/z(i)) - \beta$$

(2) if $i \in set_2$, then

$$\gamma(i) = \arctan((maxy - y(i))/z(i)) - \beta$$

(3) if $i \in set_2$, then

$$\gamma(i) = \arctan(x(i))/z(i)) - \beta(i)$$

(4) if $i \in set_4$, then

$$\gamma(i) = \arctan(y(i)/z(i)) - \beta$$

### B. Selecting and deploying redundant nodes

After adjust tilt angle $\gamma$, coverage ratio has increased significantly. There still exists many redundant sensors. Redundant sensor location in this paper is different from the existing work which concentrate on sensor deployment i.e., moving sensors to provide certain initial coverage. In our frame work, redundant sensor deployment consists of two phases: the first is to find the redundant sensors in the sensor network; the other is to relocate them to the target location.

For the first phase, we apply the set cover model algorithm to select the redundant sensors, which can be designed approximately as the followings: every time select $\Omega_{ci} \in F$ to make $|\Omega_{ci} \cap \Omega|$ is maximum to meet coverage ratio $\eta'$.

Input: coverage ratio $\eta'$ and $\Omega_{ci}$; Output: the number of sensor nodes in the process of reaching $\eta'$ and the correspondent node number.

1) Set the initial number of nodes m' to 0.
2) Select the maximum from $\{\Omega_{c1}, \Omega_{c2}, ... \Omega_{cn}\}$ to be the first selected node.
3) Add 1 to m', then deposit the new node number to the array a.
4) Remove the nodes which have been covered from set $\Omega_c$ and form a new $\Omega_c'$
5) Among the rest select the maximum one from $\Omega_{ci} \cap \Omega'$ to be the second selected node.
6) Add 1 to m', deposit the new node number to array a.
7) Repeat 4) ~ 6), until the covered node set reach $\eta'$.
8) Return to m and array a.

Then m' is the number of valid sensor nodes, (n-m') is the number of redundant sensor nodes. Then move redundant nodes according to the greedy rulers: first select the first redundant node h corresponding to the condition where $|\Omega_{ch} \cap \Omega|$ is maximum, move node h to the uncovered grid which has largest amount of neighbor grids in the range of its sensing area, then mark this newly formed coverage area as $\Omega_1$; select the second redundant node t corresponding to the condition where $|\Omega_{ct} \cap \Omega_1|$ is maximum, move this second node t to the now uncovered gird (exclude the grid covered in the first step) as the first step. Then repeat these steps until all redundant nodes are placed in the right place.

## V. EXPERIMENTAL RESULTS

### A. Simulation and Performance Evaluation

The experimental parameters are shown in Table 1. We utilize two cases (case1: N =100, case2: N=50). The simulated-experiment results are shown in Figure 6 and Figure 7. The initial are coverage ratio of monitored area G are only 32.98% and 8.23%, after adjusting tilt angle γ and after several iterations, the final coverage ratio are 93.58% and 92.18%, finally reach 99.92% and 94.88% by deploying the redundant sensor nodes, respectively.

TABLE 1. PARAMETERS IN THE EXPERIMENTS

| Parameters | Variation |
|---|---|
| Coverage ratio | 0~1 |
| Monitored area G | 500×500 units |
| Sensor number | 100 |
| α | 45° |
| β | 60° |
| z | 5~13 units |
| Maximum tilt angle K | 50° |

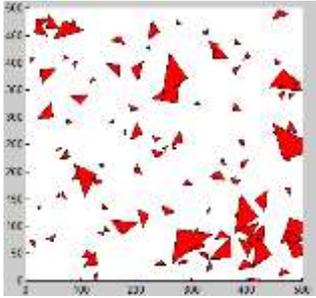

Fig7. (a) Coverage rate (32.98%)

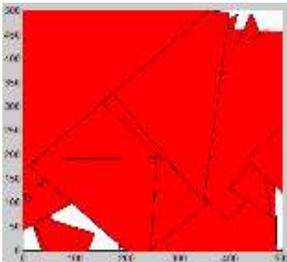 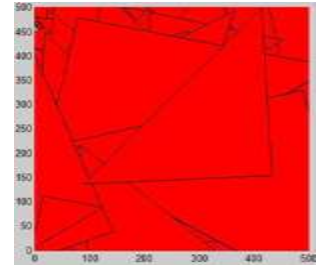

Fig7. (b) Coverage rate (93.58%)  Fig7. (c) Coverage rate (99.92%)

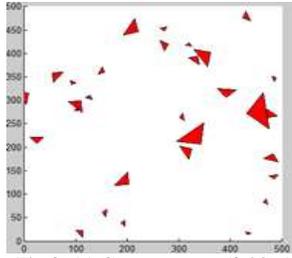

Fig.8. (a) Coverage rate (8.23%)

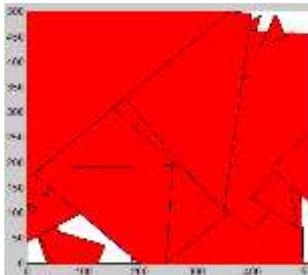 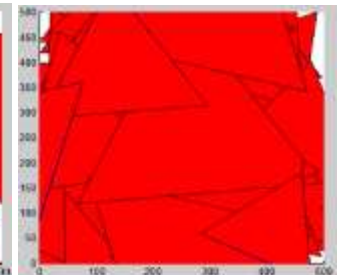

Fig.8. (b) Coverage rate (92.18%)  Fig8. (c) Coverage rate (94.88%)

## B. Convergence analysis

We conduct several experiments on the sensor scale of N=50, 80, 100 with the parameters mentioned above, respectively. Since the sensor nodes are first randomly deployed, so we repeat the experiments three times. The simulation result is shown in Table 2. After the analysis of the data in Table 2, we make a conclusion that the algorithm has good convergence on each scale, and can finish the coverage-enhancement in a period of adjustments.

TABLE 2. SENSOR NODES IMPACT ON COVERAGE RATE

| sensors | initial | final |
|---|---|---|
|  | 32.98% | 99.92% |
| 100 | 52.93% | 99.86% |
|  | 67.78% | 99.22% |
|  | 27.64% | 98.40% |
| 80 | 34.56% | 98.66% |
|  | 64.97% | 97.88% |
|  | 8.23% | 94.88% |
| 50 | 21.45% | 96.74% |
|  | 27.86% | 95.76% |

From the simulation, we know that just 50 sensor nodes with random initial deployment can realize a complete coverage. Besides, we have selected 27 sensor nodes as the redundant sensors according to the set cover algorithm when we do the experiments with 100 sensors. Compared with 28 sensors which is calculated with formula (11)、(12), the correctness of the set cover algorithm can be confirmed.

Figure 8 illustrates the improvement of 100 sensor nodes in area coverage-enhancement during the execution of the algorithm proposed in this paper (simplified as "deploy")、TDPCA proposed in the paper [11] and GA-ACE proposed in the paper [7]. It indicates the algorithm proposed in this paper can achieve faster convergence speed, which means the sensor network approach the global local optimum. In addition, redundant sensor nodes are fully taken into consideration.

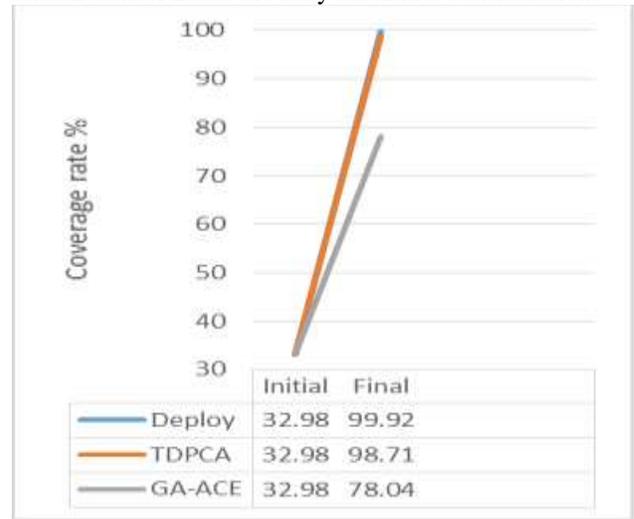

Fig.9. Coverage improvement by using different algorithm

## VI. CONCLUSION

To characterize the media sensor networks more accurately, we analyzed the 3D WMSN$_S$ model with tunable orientations and proposed a coverage-enhancing algorithm by means of deploying redundant sensor nodes to improve the efficiency of target-detecting. Through the extensive simulations experiments, we showed the effectiveness of our approach.

On-going research is take more complex environment into account, like the complex surface in 3D space and sensors can be deployed only on the surface.